# Spin Pumping in Ion-beam Sputtered $Co_2FeAl$/Mo Bilayers: Interfacial Gilbert Damping

Sajid Husain, Vineet Barwal, and Sujeet Chaudhary*

*Thin Film Laboratory, Department of Physics, Indian Institute of Technology Delhi, New Delhi 110016 (INDIA)*

Ankit Kumar, Nilamani Behera, Serkan Akansel, and Peter Svedlindh

*Ångström Laboratory, Department of Engineering Sciences, Box 534, SE-751 21 Uppsala, Sweden*

**Abstract**

The spin pumping mechanism and associated interfacial Gilbert damping are demonstrated in ion-beam sputtered $Co_2FeAl$ (CFA)/Mo bilayer thin films employing ferromagnetic resonance spectroscopy. The dependence of the net spin current transportation on Mo layer thickness, 0 to 10 nm, and the enhancement of the net effective Gilbert damping are reported. The experimental data has been analyzed using spin pumping theory in terms of spin current pumped through the ferromagnet/nonmagnetic metal interface to deduce the effective spin mixing conductance and the spin-diffusion length, which are estimated to be $1.16(\pm0.19)\times10^{19}$ m$^{-2}$ and $3.50\pm0.35$nm, respectively. The damping constant is found to be $8.4(\pm0.3)\times10^{-3}$ in the Mo(3.5nm) capped CFA(8nm) sample corresponding to a ~42% enhancement of the original Gilbert damping $(6.0(\pm0.3)\times10^{-3})$ in the uncapped CFA layer. This is further confirmed by inserting a Cu dusting layer which reduces the spin transport across the CFA/Mo interface. The Mo layer thickness dependent net spin current density is found to lie in the range of 1-3 MAm$^{-2}$, which also provides additional quantitative evidence of spin pumping in this bilayer thin film system.

*Author for correspondence: sujeetc@physics.iitd.ac.in

## I. INTRODUCTION

Magnetic damping is an exceedingly important property for spintronic devices due to its influence on power consumption and information writing in the spin-transfer torque random access memories (STT-MRAMs) [1][2]. It is therefore of high importance to study the generation, manipulation, and detection of the flow of spin angular momentum to enable the design of efficient spin-based magnetic memories and logic devices [3]. The transfer of spin angular momentum known as spin pumping in ferromagnetic (FM)/ nonmagnetic (NM) bilayers provides information of how the precession of the magnetization transfers spin angular momentum into the adjacent nonmagnetic metallic layer [4]. This transfer (pumping) of spin angular momentum slows down the precession and leads to an enhancement of the effective Gilbert damping constant in FM/NM bilayers. This enhancement has been an area of intensive research since the novel mechanism (theory) of spin pumping was proposed by Arne Brataas *et al.* [5] [6]. The amount of spin pumping is quantified by the magnitude of the spin current density at the FM/NM interface and theoretically [7] described as $\mathbf{J}_S^{eff} = \frac{\hbar}{4\pi} g_{eff}^{\uparrow\downarrow} \left( \mathbf{m} \times \frac{d\mathbf{m}}{dt} \right)$ where **m** is the magnetization unit vector, $\mathbf{J}_S^{eff}$ is the effective spin current density pumped into the NM layer from the FM layer (portrayed in Fig. 1), and $g_{eff}^{\uparrow\downarrow}$ is the spin mixing conductance which is determined by the reflection coefficients of conductance channels at FM/NM interface [5].

To date, a number of NM metals, such as Pt, Au, [5], Pd [8][9], β-Ta [10] and Ru [11], *etc.* have been extensively investigated with regards to their performance as spin sink material when in contact with a FM. It is to be noted here that none of the Pt, Pd, Ru, and Au is an earth abundant material [12]. Thus, there is a natural need to search for new non-magnetic materials which could generate large spin current at the FM/NM interface. In this study, we have explored the potential of the transition metal molybdenum (Mo) as a new candidate material for spin pumping owing to the fact that Mo possesses a large spin-orbit coupling [13]. To the best of our knowledge, Mo has not been used till date for the study of spin pumping effect in a FM/NM bilayer system.

In a FM/NM bilayer and/or multilayer systems, there are several mechanisms for dissipation of the spin angular momentum which are categorized as intrinsic and extrinsic. In the intrinsic category, the magnon-electron coupling, *i.e.*, spin-orbit coupling (SOC) contributes significantly [14]. Among the extrinsic category, the two-magnon scattering (TMS) mechanism is linked to the inhomogeneity and interface/surface roughness of the heterostructure, *etc.* [15] [16] [17]. For large SOC, interfacial *d-d* hybridization between the NM and FM layers is highly desirable [16]. Thus, the FM-NM interfacial hybridization is expected to result in enhancement of the transfer of spin angular momentum from the FM to the NM layer, and hence the NM layer can act as a spin reservoir (sink) [18]. But, the NM metallic layer does not always act as a perfect spin reservoir due to the spin accumulation effect which prevents transfer of angular momentum to some extent and as a result, a backflow of spin-current towards the FM [6] is estabished. While the flow of spin angular momentum through the FM/NM interface is determined by the effective spin-mixing conductance ($g_{eff}^{\uparrow\downarrow}$) at the interface, the spin backflow is governed by the spin diffusion length ($\lambda_d$). It is emphasized here that these parameters ($g_{eff}^{\uparrow\downarrow}$ and $\lambda_d$) are primarily tuned by appropriate selection of a suitable NM layer.

In this work, we have performed ferromagnetic resonance (FMR) measurements to explore the spin pumping phenomenon and associated interfacial Gilbert damping enhancement in the Co$_2$FeAl(8nm)/Mo($t_{Mo}$) bilayer system, $t_{Mo}$ is the thickness of Mo, which is varied from 0 to 10 nm. The $t_{Mo}$ dependent net spin current transfer across the interface and spin diffusion length of Mo are estimated. The choice of employing the Heusler alloy CoFe$_2$Al (CFA) as a thin FM layer lies in its half metallic character anticipated at room temperature [19] [20], a trait which is highly desirable in any spintronic device operating at room temperature.

## II. EXPERIMENTAL DETAILS

The CFA thin films with fixed thickness of 8 nm were grown on naturally oxidized Si(100) substrate at 573K temperature using an ion-beam sputtering deposition system (NORDIKO-3450). The substrate temperature (573K) has been selected following the growth optimization reported in our previous reports [21] [20] [22]. On the top of the CFA layer, a Mo film with thickness $t_{Mo}$ ($t_{Mo}$ =0, 0.5, 1.0, 1.5, 2.0, 3.0, 4, 5, 7, 8 and 10 nm) was deposited *in situ* at room temperature. In addition, a trilayer structure of CFA(8)/Cu(1)/Mo(5) was also prepared to understand and confirm the effect of an additional interface on the Gilbert damping (spin pumping). Numbers in parenthesis are film thicknesses in *nm*. All the samples were prepared at a constant working pressure of ~8.5×10$^{-5}$ Torr (base vacuum ~ 1.0×10$^{-7}$ Torr); Ar gas was directly fed at 4 sccm into the *rf*-ion source operated at 75W with the deposition rates of 0.03nm/s and 0.02nm/s for CFA and Mo, respectively. The deposition rate for Cu was 0.07nm/s at 80 W. The samples were then cut to 1×4 mm$^2$ to record the FMR spectra employing a homebuilt FMR set-up [21] [23]. The data was collected in DC-magnetic field sweep mode by keeping the microwave frequency fixed. The saturation magnetization was measured using the Quantum Design make Physical Property Measurement System (Model PPMS *Evercool-II*) with the vibrating sample magnetometer option (QD PPMS-VSM). The film density, thickness and interface/surface roughness were estimated by simulating the specular X-ray reflectivity (XRR) spectra using the PANalytical X'Pert reflectivity software (Ver. 1.2 with segmented fit). To determine surface morphology/microstructure (e.g., roughness), topographical imaging was performed using the 'Bruker dimension ICON *scan assist*' atomic force microscope (AFM). All measurements were performed at room temperature.

### III. RESULTS AND DISCUSSIONS

#### A. X-ray Reflectivity and Atomic Force Microscopy: Interface/surface analysis

Figure 2 shows the specular XRR spectra recorded on all the CFA(8)/Mo($t_{Mo}$) bilayer thin films. The fitting parameters were accurately determined by simulating (red lines) the experimental curves (filled circles) and are presented in Table-I. It is evident that for the smallest NM layer thickness, Mo(0.5nm), the estimated values of the roughness from XRR and AFM are slightly larger in comparison to the thickness of the Mo layer which indicates that the surface coverage of Mo layer is not enough to cover all of the CFA surface in the CFA(8)/Mo(0.5) bilayer sample (modeled in Fig. 3(a)). For $t_{Mo} \geq$ 1nm, the film roughness is smaller than the thickness (indicating that the Mo layer coverage is uniform as modeled in Figs. 3(b)-(c)). For the thicker layers of Mo ($t_{Mo} \geq$ 5nm) the estimated values of the surface roughness as estimated from both XRR and AFM are found to be similar ~0.6nm (c.f. the lowest right panel in Fig. 2).

#### B. Ferromagnetic Resonance Study

The FMR spectra were recorded on all samples in 5 to 11 GHz range of microwave frequencies. Figure. 4(a) shows the FMR spectra recorded on the CFA(8)/Mo(5) bilayer thin film. The FMR spectra ($I_{FMR}$) were fitted with the derivative of symmetric and anti-symmetric Lorentzian functions to extract the line-shape parameters, *i.e.*, resonant field $H_r$ and linewidth $\Delta H$, given by [24] [21]:

$$I_{FMR} = \frac{\partial(U)}{\partial H} = S\frac{\partial\left(F_S(H_{ext})\right)}{\partial H_{ext}} + A\frac{\partial\left(F_A(H_{ext})\right)}{\partial H_{ext}}$$

$$= -S\frac{\left(\frac{\Delta H}{2}\right)^2 (H_{dc} - H_r)}{\left[(H_{dc} - H_r)^2 + \left(\frac{\Delta H}{2}\right)^2\right]^2} + A\frac{\frac{\Delta H}{2}\left[\left(\frac{\Delta H}{2}\right)^2 - (H_{dc} - H_r)^2\right]}{\left[(H_{dc} - H_r)^2 + \left(\frac{\Delta H}{2}\right)^2\right]^2}, \quad (1)$$

where $F_S(H_{ext})$ and $F_A(H_{ext})$ are the symmetric and anti-symmetric Lorentzian functions, respectively, with $S$ and $A$ being the corresponding coefficients. Symbol 'U' refers to the raw signal voltage from the VNA. The linewidth $\Delta H$ is the full width at half maximum (FWHM), and $H_{dc}$ is the applied DC-magnetic field.

The $f$ vs. $\mu_0 H_r$ plots are shown in Fig. 4(b). These are fitted using the Kittel's formula [25]:

$$f = \frac{\mu_0 \gamma}{2\pi}\sqrt{(H_r + H_K)(H_r + H_K + M_{eff})}, \quad (2)$$

where $\gamma$ is the gyromagnetic ratio; $\gamma = g\mu_B / \hbar$ (1.76×10[11]s[-1]T[-1]) with $g$ being the Lande's splitting factor; taken as 2, $\mu_0 M_{eff}$ is the effective saturation magnetization, and $\mu_0 H_K$ is the uniaxial anisotropy field. The values of $\mu_0 M_{eff}$ are comparable to the values of $\mu_0 M_S$ (obtained from VSM measurements) as is shown in Fig. 4(c). Figure 4(e) shows the variation of $\mu_0 H_K$ with $t_{Mo}$ from which the decrease in $\mu_0 H_K$ with increase in $t_{Mo}$ is clearly evident. This observed reduction in $\mu_0 H_K$ could possibly stem from the spin accumulation increasing with increasing $t_{Mo}$ [26]. The FMR spectra was also recorded on CFA(8)/Cu(1)/Mo(5) trilayer thin film for the comparison with the results of CFA(8)/Mo(5) bilayer. The magnitudes of $\mu_0 M_{eff}$ ($\mu_0 H_k$) for CFA(8)/Mo(5) and CFA(8)/Cu(1)/Mo(5) are found to be 1.33±0.08T (0.55±0.15mT) and

1.30±0.04T (3.21±0.13 mT), respectively. Further, Fig. 4(d) shows the $\mu_0 H_r$ vs. $t_{Mo}$ behavior at different constant frequencies ranging between 5 to 11 GHz. The observed values of $\mu_0 H_r$ are constant for all the Mo capped layers which clearly indicates that the dominant contribution to the observed resonance spectra arises from the intrinsic effect, *i.e.*, magnon-electron scattering [27].

### C. Mo thickness-dependent spin pumping

Figure 5(a) shows the linewidth $\mu_0 \Delta H$ vs. $f$ (for clarity, the results are shown only for a few selected film samples). The frequency dependent linewidth can mainly have two contributions; the intrinsic magnon-electron scattering contribution, and the extrinsic two-magnon scattering (TMS) contribution. The extrinsic TMS contribution in linewidth has been analysed (not presented here) using the methods given by Arias and Mills [28]. A similar analysis was reported in one of our previous studies on the CFA/Ta system [21]. For the present case, the linewidth analysis shows that inclusion of the TMS part does not affect the Gilbert damping, which means the TMS contribution is negligible in our case. Now, the effective Gilbert damping constant $\alpha_{eff}$ can be estimated using,

$$\Delta H = \Delta H_0 + \frac{4\pi \alpha_{eff} f}{\mu_0 \gamma}. \quad (2)$$

Here, $\Delta H_0$ is the frequency independent contribution from sample inhomogeneity, while the second term corresponds to the frequency dependent contribution associated with the intrinsic Gilbert relaxation. Here, $\alpha_{eff}$, defined as $\alpha_{eff} = \alpha_{SP} + \alpha_{CFA}$, is the effective Gilbert damping which includes the intrinsic value of CFA ($\alpha_{eff}$) and a spin pumping contribution ($\alpha_{SP}$) from the CFA/Mo bilayer.

The extracted effective Gilbert damping constant values for different $t_{Mo}$ are shown in Fig 5(b). An enhancement of the Gilbert damping constant with the increase of the Mo layer thickness is clearly observed, which is anticipated owing to the transfer of spin angular momenta by spin pumping from CFA to the Mo layer at the CFA/Mo interface. The value of $\alpha_{eff}$ is found to increase up to $8.4(\pm 0.3) \times 10^{-3}$ with the increase in $t_{Mo}$ ($\geq 3.5$nm), which corresponds to ~42% enhancement of the damping constant due to spin pumping. It is remarkable that such a large change in Gilbert damping is observed for the CFA/Mo bilayer; the change is comparable to those reported when a high SOC NM such as Pt [8], Pd [29] [9], Ru [11], and Ta [30] is employed in FM/NM bilayers. Here, we would like to mention that the enhancement of the Gilbert damping can, in principle, also be explained by extrinsic two-magnon scattering (TMS) contributions in CFA/Mo($t_{Mo}$) bilayers by considering the variation of $H_r$ with NM thickness [27]. In our case, the $\mu_0 H_r$ is constant for all $t_{Mo}$ (c.f. Fig. 4(d)). Thus the extrinsic contribution induced increase in $\alpha_{eff}$ is negligibly small and hence the enhancement of the damping is dominated by the spin pumping mechanism. The estimated values of $\mu_0 \Delta H_0$ are found to vary from 0.6 to 2.5 mT in the CFA/Mo($t_{Mo}$) thin films. The variation in $\mu_0 \Delta H_0$ is assigned to the finite, but small, statistical variations in sputtering conditions between samples with different $t_{Mo}$.

Further, to affirm the spin pumping in the CFA/Mo bilayer system, a copper (Cu) dusting layer was inserted at the CFA/Mo interface. Figure 5(c) compares the linewidth *vs. f* plot of the

CFA(8)/Mo(5) and CFA(8)/Cu(1)/Mo(5) heterostructures. The Gilbert damping was found to decrease from 8.4(±0.3)×10$^{-3}$ to 6.4(±0.3)×10$^{-3}$ after inserting the Cu(1) thin layer, which is comparable to the value of the uncapped CFA(3.5) sample. It may be noted that Cu has a very large spin diffusion length ($\lambda_d$ ~300nm) but weak SOC strength [32]. Due to the weak SOC, the asymmetry in the band structure at the FM/Cu interface would thus lead to a non-equilibrium spin accumulation at the CFA/Cu interface [33]. This spin accumulation opposes the transfer of angular momentum into the Mo layer and hence the Gilbert damping value, after insertion of the dusting layer, is found very similar to that of the single layer CFA film. It is also known that enhancement of damping in the FM layer (when coupled to the NM layer) can occur due to the magnetic proximity effect [34]. However, we did not find any evidence in favor of the proximity effect as the effective saturation magnetization did not show any increase on the insertion of the ultrathin Cu dusting layer at CFA/Mo bilayer interface, which supports our claim of absence of spin pumping in the CFA/Cu/Mo trilayer sample.

The flow of angular momentum across the FM/NM bilayer interface is determined by the effective complex spin-mixing conductance $g_{eff}^{\uparrow\downarrow} = \text{Re}(g_{eff}^{\uparrow\downarrow}) + i\,\text{Im}(g_{eff}^{\uparrow\downarrow})$, defined as the flow of angular momentum per unit area through the FM/NM metal interface created by the precessing moments in the FM layer. The term *effective* spin-mixing conductance is being used because it contains the forward and backflow of spin momentum at the FM/NM interface. The imaginary part of the spin-mixing conductance is usually assumed to be negligibly small $\left[\text{Re}(g_{eff}^{\uparrow\downarrow}) \gg \text{Im}(g_{eff}^{\uparrow\downarrow})\right]$ as compared to the real part [35] [36], and therefore, to determine the real part of the spin-mixing conductance, the obtained $t_{Mo}$ dependent Gilbert damping is fitted with the relation [29],

$$\alpha_{eff} = \alpha_{CFA} + \text{Re}(g_{eff}^{\uparrow\downarrow})\frac{g\mu_B}{4\pi M_S}\frac{1}{t_{CFA}}\left(1 - e^{\frac{-2t_{Mo}}{\lambda_d}}\right), \quad (3)$$

where $\alpha_{CFA}$ is the damping for a single layer CFA without Mo capping layer, $\text{Re}(g_{eff}^{\uparrow\downarrow})$ is given in units of m$^{-2}$, $\mu_B$ is the Bohr magneton, and $t_{CFA}$ is a CFA layer thickness. The exponential term describes the reflection of spin-current from Mo/air interface. Figure 5(b) shows the variation of the effective Gilbert damping constant with $t_{Mo}$ and the fit using Eqn. (3) (red line). The values of $\text{Re}(g_{eff}^{\uparrow\downarrow})$ and $\lambda_d$ are found to be 1.16($\pm$0.19)$\times 10^{19}$ m$^{-2}$ and 3.5$\pm$0.35 nm, respectively. The value of the spin-mixing conductance is comparable to those recently reported in FM/Pt(Pd) thin films such as Co/Pt (1-4 $\times 10^{19}$ m$^{-2}$) [8] [33], YIG/Pt (9.7 $\times 10^{18}$ m$^{-2}$) [37], Fe/Pd (1$\times 10^{20}$ m$^{-2}$) [9], and Py/Pd(Pt) (1.4(3.2) $\times 10^{18}$ m$^{-2}$) [34].

We now calculate the net intrinsic interfacial spin mixing conductance $G^{\uparrow\downarrow}$ which depends on the thickness and the nature of the NM layer as per the relation [9] [38],

$$G^{\uparrow\downarrow}(t_{Mo}) = \text{Re}(g_{eff}^{\uparrow\downarrow})\left[1 + \left[\sqrt{\frac{4\varepsilon}{3}}\tanh\frac{t_{Mo}}{\lambda_d}\right]^{-1}\right]^{-1}, \quad (4)$$

where $\varepsilon = (Ze^2/\hbar c)^4$ is a material dependent parameter (Z is the atomic number of Mo *i.e.*, 42 and c is the speed of light) whose value for Mo is 0.0088. Using Eq. (4), $G^{\uparrow\downarrow}(t_{Mo})$ values have been computed for various $t_{Mo}$; the results are shown in Fig. 6(a). The $t_{Mo}$ dependence of $G^{\uparrow\downarrow}$ clearly suggests that the spin mixing conductance critically depends on the NM layer properties. For bilayers with $t_{Mo} \geq 6$ nm, $G^{\uparrow\downarrow}$ attains its saturation value, which is quite comparable with those reported for Pd and Pt [34] [37]. Understandably, such a large value of the spin mixing conductance will yield a large spin current into the adjacent NM layer [6] [7] [37] [33]. In the next section, we have estimated the spin current from the experimental FMR data and discuss the same with regards to spin pumping in further detail.

### D. Spin current generation in Mo due to spin pumping

The enhancement of the Gilbert damping observed in the CFA(8)/Mo($t_{Mo}$) bilayers (Fig. 5(b)) is generally interpreted in terms of the spin-current generated in Mo layer by the spin pumping mechanism at the bilayer interface (Fig. 1). The associated net effective spin current density in Mo is described by the relation [38] [39]:

$$J_S^{eff}(t_{Mo}) = G^{\uparrow\downarrow}(t_{Mo}) \frac{2e}{\hbar} \frac{\gamma^2 \hbar \left(\mu_0 h_{rf}\right)^2}{8\pi \alpha_{eff}^2} \left[ \frac{\mu_0 \gamma M_{eff} + \sqrt{\left(\mu_0 \gamma M_{eff}\right)^2 + 4\omega^2}}{\left(\mu_0 \gamma M_{eff}\right)^2 + 4\omega^2} \right], \quad (5)$$

where $\omega = 2\pi f$ and $h_{rf}$ is the *rf*-field (26 A/m) in the strip-line of our co-planar waveguide. $G^{\uparrow\downarrow}(t_{Mo})$ is the net intrinsic interfacial spin mixing conductance discussed in the previous section (Fig. 6). The estimated values of $J_S^{eff}(t_{Mo})$ for different microwave frequencies are shown in Fig. 7. It is clearly observed that the spin current density increases with the increase in $t_{Mo}$, the increase becomes relatively less at higher $t_{Mo}$, which indicates the progressive spin current generation in Mo. Such an appreciable change in current density directly provides evidence of the interfacial enhancement of the Gilbert damping in these CFA/Mo bilayers.

Further, it would be interesting to investigate the effect on the spin current generation in Mo layer if an ultrathin dusting layer of Cu is inserted at the CFA/Mo interface. In principle, on insertion of a thin Cu layer, the spin pumping should cease because of the unmatched band structure between the CFA/Cu and Cu/Mo interfaces owing to the insignificant SOC in Cu. This is in consonance with the observed decrease in Gilbert damping back to the value for the uncapped CFA layer (c.f. Fig. 5(c) and associated discussion). The spin-mixing conductance of the trilayer heterostructure can be evaluated by $\Delta\alpha_{eff} = g\mu_B g_{eff}^{\uparrow\downarrow}/\mu_0 M_S t_{CFA}$ [29], where $\Delta\alpha_{sp} = \alpha_{eff} - \alpha_{CFA}$ is the spin-pumping induced Gilbert damping contribution which for the CFA/Cu/Mo trilayer is quite small, *i.e.*, $4.0(\pm0.3)\times10^{-4}$ after Cu insertion. For the trilayer, $g_{eff}^{\uparrow\downarrow}$ is found to be $1.49(\pm0.12)\times10^{17}$ m$^{-2}$ which is two orders of magnitude smaller compared to that of the CFA/Mo bilayers. Furthermore, using the values of $g_{eff}^{\uparrow\downarrow}$, $\mu_0 M_{eff}$ and $\alpha_{eff}$ for the CFA/Cu/Mo trilayer heterostructure in Eqn. (5) and for $f$ = 9GHz, the spin current density is found be $0.0278(\pm0.0013)$ MA/m$^2$, which is two order of magnitude smaller than that in the CFA/Mo bilayers. Thus, the reduction in $\alpha_{eff}$ and $J_S^{eff}$ subsequent to Cu dusting is quite comparable to previously reported results [33] [40].

## IV. CONCLUSIONS

We have systematically investigated the changes in the spin dynamics in the ion-beam sputtered Co$_2$FeAl (CFA)/Mo($t_{Mo}$) bilayers for various $t_{Mo}$ at constant CFA thickness of 8nm. Increasing the Mo layer thickness to its spin diffusion length; CFA(8)/Mo($t_{Mo} = \lambda_d$), the effective Gilbert damping constant increases to $8.4(\pm0.3)\times10^{-3}$ which corresponds to about ~42% enhancement with respect to the $\alpha_{eff}$ value of $6.0(\pm0.3)\times10^{-3}$ for the uncapped CFA layer (*i.e.*, without the top Mo layer). We interpret our results based on the spin-pumping effects wherein the effective spin-mixing conductance, and spin-diffusion length are found to be $1.16(\pm0.19)\times10^{19}$ m$^{-2}$ and $3.50\pm0.35$nm, respectively. The spin pumping is further confirmed by inserting an ultrathin Cu layer at the CFA/Mo interface. The overall effect of the damping constant enhancement observed when Mo is deposited over CFA is remarkably comparable to the far less-abundant non-magnetic metals that are currently being used for spin pumping applications. From this view point, the demonstration of the new material, *i.e.*, Mo, as a suitable spin pumping medium is indispensable for the development of novel STT spintronic devices.

## ACKNOWLEDGMENTS

One of the authors SH acknowledges the Department of Science and Technology, Govt. of India for providing the INSPIRE Fellowship. Authors thank the NRF facilities of IIT Delhi for AFM imaging. This work was in part supported by Knut and Alice Wallenberg (KAW) Foundation Grant No. KAW 2012.0031. We also acknowledge the Ministry of Information Technology, Government of India for providing the financial grant.

Table: 1 Summary of XRR simulated parameters, *i.e.*, $\rho$, $t_{FM}$, $t_{Mo}$, and σ for the bilayer thin films [Si/CFA(8)/Mo($t_{Mo}$)]. Here $\rho$, $t_{FM}$, $t_{Mo}$, and σ refer to the density, thickness, and interface width of the individual layers, respectively.

| | CFA (Nominal thickness = 8nm) | | | Mo | | | MoOx | | |
|---|---|---|---|---|---|---|---|---|---|
| S.No. | ρ(gm/cc)±0.06 | $t_{FM}$(nm)±0.01 | σ(nm) ±0.03 | ρ(gm/cc)±0.05 | $t_{Mo}$(nm)±0.01 | σ(nm) ±0.03 | ρ(gm/cc)±0.06 | $t$(nm)±0.01 | σ(nm) ±0.03 |
| 1 | 7.35 | 7.00 | 0.20 | 6.05 | 0.58 | 0.94 | 4.07 | 0.97 | 0.59 |
| 2 | 7.31 | 8.17 | 0.35 | 8.58 | 1.00 | 0.54 | 4.04 | 0.82 | 0.35 |
| 3 | 7.50 | 7.22 | 0.80 | 10.50 | 1.50 | 0.52 | 5.00 | 1.08 | 0.5 |
| 4 | 7.50 | 8.18 | 0.37 | 9.94 | 2.00 | 0.64 | 4.38 | 0.85 | 0.45 |
| 5 | 7.00 | 7.00 | 1.00 | 9.50 | 3.00 | 0.60 | 5.17 | 1.00 | 0.37 |
| 6 | 7.00 | 8.28 | 0.56 | 10.45 | 3.46 | 0.78 | 4.38 | 1.01 | 0.4 |
| 7 | 7.29 | 8.00 | 0.44 | 9.43 | 4.86 | 0.26 | 6.50 | 0.98 | 0.56 |
| 8 | 7.22 | 7.79 | 0.98 | 10.50 | 6.47 | 0.67 | 4.81 | 1.03 | 0.62 |
| 9 | 7.00 | 8.12 | 0.15 | 9.29 | 8.21 | 0.64 | 4.00 | 0.96 | 0.8 |
| 10 | 7.64 | 8.00 | 0.17 | 9.23 | 10.26 | 0.67 | 5.00 | 1.17 | 0.73 |

**Figure captions**

FIG. 1. (color online) Schematic of the CFA/Mo bilayer structure used in our work portrayed for an example of spin current density $J_S^{eff}$ generated at the CFA/Mo interface by spin pumping.

FIG. 2 XRR spectra and the AFM topographical images of Si/CFA(8)/Mo($t_{Mo}$). In the respective XRR spectra, circles represent the recorded experimental data points, and lines represent the simulated profiles. The estimated values of the surface roughness in the entire sample series as obtained from XRR and AFM topographical measurements are compared in the lowest right panel. The simulated parameters are presented in the Table-I. All AFM images were recorded on a scan area of $10 \times 10 \mu m^2$.

FIG.3: The atomic representation (model) of the growth of the Mo layer (yellow sphere) on top of the CFA (blue spheres) layer. The film changes from discontinuous to continuous as the thickness of the Mo layer is increased. Shown are the 3 different growth stages of the films: (a) least coverage (b) partial coverage and (c) full coverage.

FIG. 4: (a) Typical FMR spectra recorded at various frequencies (numbers in graph are the microwave frequencies in GHz) for the Si/SiO$_2$/CFA(8)/Mo(5) bilayer sample (symbols correspond to experimental data and red lines are fits to the Eqn. (1)) Inset: FMR spectra of CFA single layer (filled circles) and CFA(8)/Mo(2) bilayer (open circles) samples measured at 5GHz showing the increase in linewidth due to spin pumping. (b) The resonance field $\mu_0 H_r$ vs. $f$ for all the samples (red lines are the fits to the Eqn. (2). (c) Effective magnetization (scale on left) and saturation magnetization (scale on right) vs. $t_{Mo}$. The solid line represents the bulk value of the saturation magnetization of Co$_2$FeAl. (d) The resonance field $\mu_0 H_r$ vs. $t_{Mo}$ at different constant

frequencies for CFA(8)/Mo($t_{Mo}$) bilayer thin films. (e) Anisotropy field $\mu_0 H_K$ vs. $t_{Mo}$. (f) Comparison of $\mu_0 H_r$ vs. $f$ for the CFA(8)/Mo(5) and CFA(8)/Cu(1)/Mo(5) samples.

FIG. 5: (a) Linewidth vs. frequency for Si/SiO$_2$/CFA(8)/Mo($t_{Mo}$) bilayer thin films. (b) Effective Gilbert damping constant vs. Mo layer thicknesses. (c) $\mu_0 \Delta H$ vs. $f$ for CFA(8)/ Mo(5) and CFA(8)/Cu(1)/Mo(5) films.

FIG. 6: Intrinsic spin-mixing conductance vs. $t_{Mo}$ of the CFA(8)/Mo($t_{Mo}$) bilayers.

FIG. 7. The effective spin current density (generated in Mo) vs. $t_{Mo}$ at different microwave frequencies calculated using Eqn. (5)

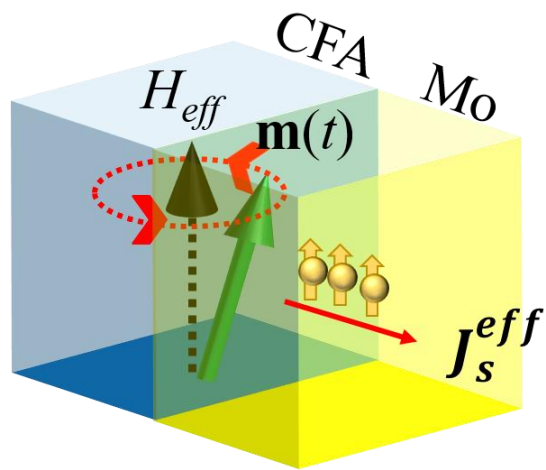

FIG. 1

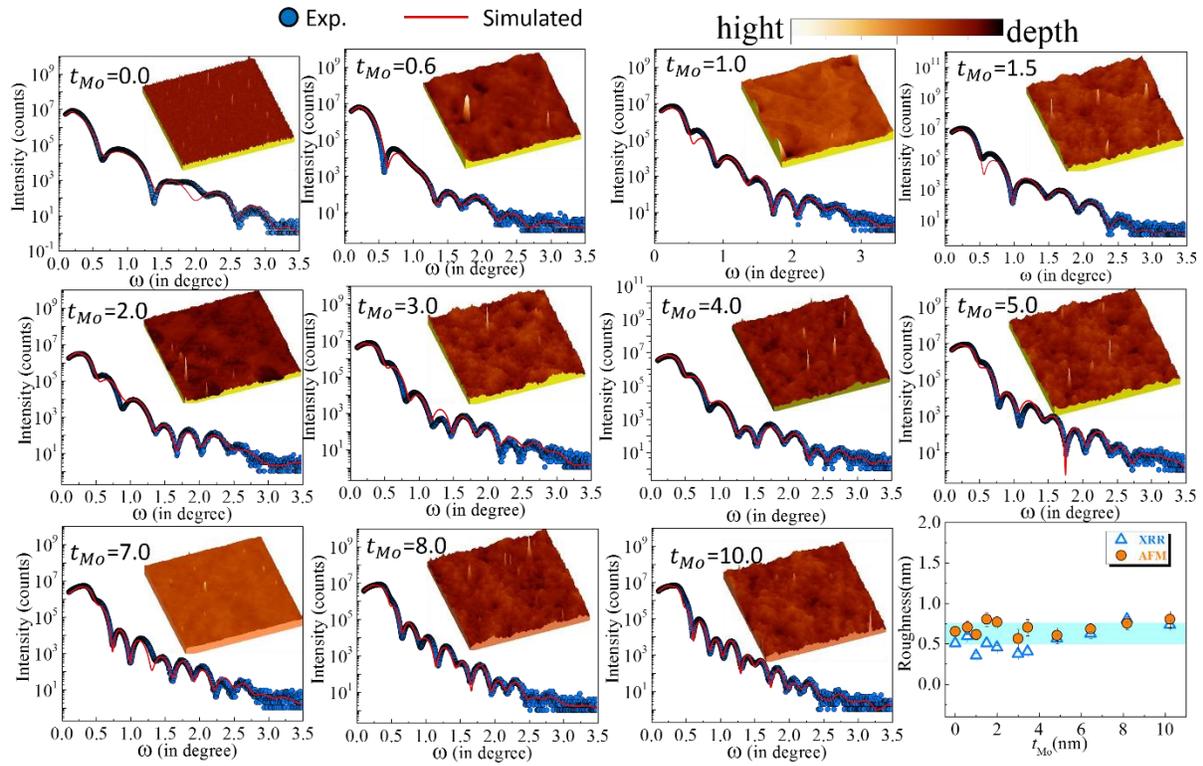

FIG. 2

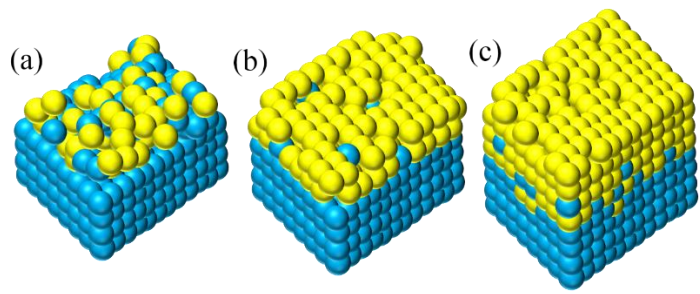

FIG. 3

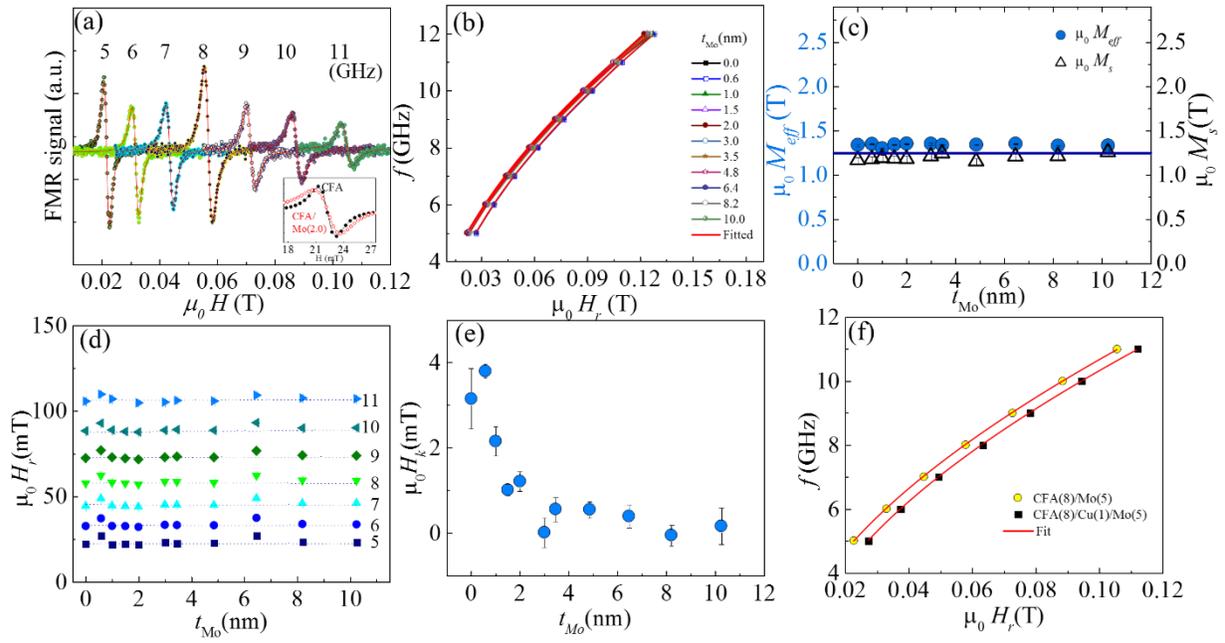

FIG. 4

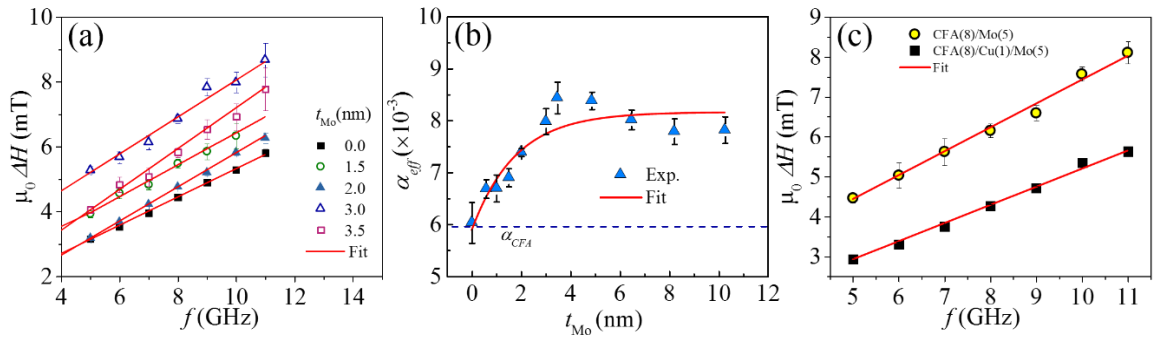

FIG. 5

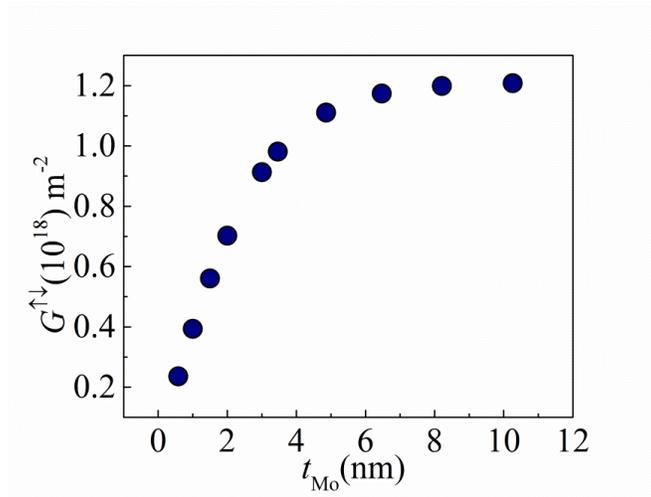

FIG. 6

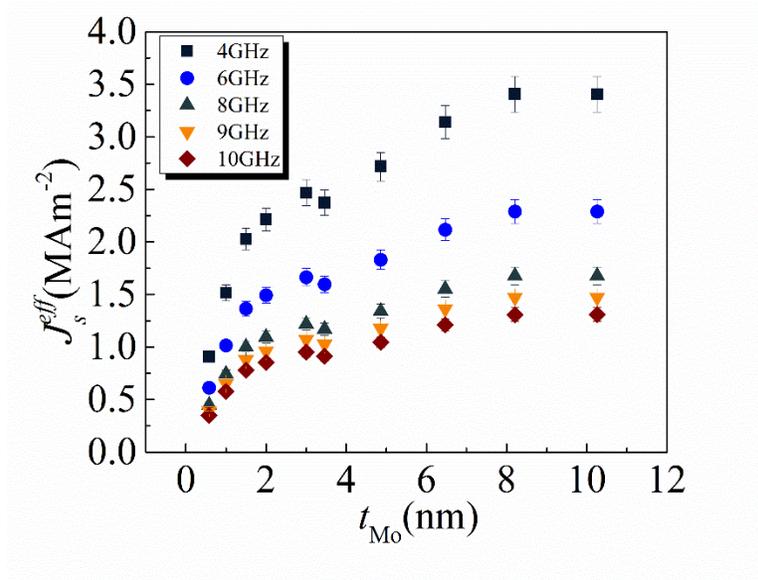

FIG. 7